\documentclass[sigconf]{acmart}

\usepackage{subcaption}

\AtBeginDocument{%
  }

\copyrightyear{2025}
\acmYear{2025}
\setcopyright{cc}
\setcctype{by}
\acmConference[Koli Calling '25]{25th Koli Calling International Conference on Computing Education Research}{November 11--16, 2025}{Koli, Finland}
\acmBooktitle{25th Koli Calling International Conference on Computing Education Research (Koli Calling '25), November 11--16, 2025, Koli, Finland}
\acmPrice{}
\acmDOI{10.1145/3769994.3769997}
\acmISBN{979-8-4007-1599-0/25/11}

\begin{document}

\title{Designing for Novice Debuggers: A Pilot Study on an AI-Assisted Debugging Tool}

\author{Oka Kurniawan}
\email{oka_kurniawan@sutd.edu.sg}
\orcid{1234-5678-9012}

\affiliation{%
  \institution{Singapore University of Technology and Design}
  \city{Singapore}
  \country{Singapore}
}
\author{Erick Chandra}
\email{erick_chandra@sutd.edu.sg}
\orcid{0009-0005-9558-1200}
\affiliation{%
  \institution{Singapore University of Technology and Design}
  \city{Singapore}
  \country{Singapore}
}

\author{Christopher M. Poskitt}
\orcid{0000-0002-9376-2471}
\email{cposkitt@smu.edu.sg}
\affiliation{%
  \institution{Singapore Management University}
  \city{Singapore}
  \country{Singapore}
}

\author{Yannic Noller}
\email{yannic.noller@acm.org}
\orcid{0000-0002-9318-8027}
\affiliation{%
  \institution{Ruhr University Bochum}
  \city{Bochum}
  \country{Germany}
}

\author{Kenny Tsu Wei Choo}
\email{kennytwchoo@gmail.com}
\orcid{0000-0003-3845-9143}
\affiliation{%
  \institution{Singapore University of Technology and Design}
  \city{Singapore}
  \country{Singapore}
}

\author{Cyrille Jegourel}
\email{cyrille_jegourel@sutd.edu.sg}
\orcid{0000-0003-2770-8394}
\affiliation{%
  \institution{Singapore University of Technology and Design}
  \city{Singapore}
  \country{Singapore}
}

\begin{abstract}
 Debugging is a fundamental skill that novice programmers must develop. Numerous tools have been created to assist novice programmers in this process. Recently, large language models (LLMs) have been integrated with automated program repair techniques to generate fixes for students' buggy code. However, many of these tools foster an over-reliance on AI and do not actively engage students in the debugging process. In this work, we aim to design an intuitive debugging assistant, \textsc{CodeHinter}, that combines traditional debugging tools with LLM-based techniques to help novice debuggers fix semantic errors while promoting active engagement in the debugging process. We present findings from our second design iteration, which we tested with a group of undergraduate students. Our results indicate that the students found the tool highly effective in resolving semantic errors and significantly easier to use than the first version. Consistent with our previous study, error localization was the most valuable feature. Finally, we conclude that any AI-assisted debugging approach should be personalized based on user profiles to optimize their interactions with the tool. 
\end{abstract}



\maketitle

\section{Introduction}

The first hurdle novice programmers encounter when writing code is ensuring that it is free of syntax errors \cite{Smith2019}. Although most compilers and interpreters generate error messages that provide information about these errors, many novice programmers struggle to interpret and fix them \cite{Denny2012}. To address these challenges, numerous tools and approaches have been developed to help novice programmers read, interpret, and resolve error messages effectively \cite{Campbell2014, Leinonen2023}. However, many of these tools primarily focus on improving error messages rather than guiding students through the debugging process \cite{Becker2019}.

Recent advances in AI-assisted programming have introduced large language model (LLM) tools as potential debugging assistants, capable of automatically identifying and fixing errors.
LLMs have already shown strong accuracy in correcting syntax errors, opening up new possibilities for automated debugging support~\cite{Zhang2023, Joshi2023}.
Moreover, LLM-based tools such as ChatGPT and GitHub Copilot can provide explanations of syntax errors along with code examples that help learners understand and correct their mistakes~\cite{chatgpt, copilot}. 

While LLM-based tools provide valuable assistance in fixing syntax errors, their ability to help novice programmers identify, understand, and resolve semantic errors remains limited. First, studies have shown that while LLM-based tools can resolve some semantic errors, their accuracy varies depending on the complexity of the problem~\cite{Zhang2023}. Second, the generated solutions may differ significantly from students' original buggy code, making it difficult for novice programmers to understand how to modify their code accordingly~\cite{Zhang2024}. Third, LLM-based tools may also introduce new errors, which novice programmers often struggle to recognize or correct, leading to a negative learning experience.

Finally, and most importantly, many LLM-based tools generate complete solutions to semantic errors rather than guiding users through the step-by-step debugging process that is essential for learning. As a result, novice programmers may develop an over-reliance on AI tools, which can hinder their ability to debug independently~\cite{Collins2023}. This dependence not only affects the accuracy of solutions but also impairs the ability of students to develop critical problem-solving skills. Without structured guidance, students may adopt AI-generated solutions without fully understanding how to troubleshoot and resolve issues on their own. Proficiency in testing, debugging, and fixing code is a fundamental skill for computer science (CS) graduates, and its importance is even greater in the era of AI-assisted programming, where AI tools enhance productivity but cannot substitute for a programmer’s ability to independently reason through errors.

Our work introduces \textsc{CodeHinter}, an interactive debugging tool that goes beyond automatic error correction to actively guide students through diagnosing and fixing their own code. By integrating fault localization techniques with LLM-powered hints directly into the IDE through a simple one-click interface, \textsc{CodeHinter} promotes step-by-step problem solving, deepens students’ understanding of errors, and reduces reliance on fully automated solutions. We evaluate its usability through a user study, highlighting how structured, interactive guidance can strengthen debugging skills and improve programming education.

\section{Related Work}

In this section, we review related work on how LLMs assist novice programmers in fixing errors. We examine various tools designed to help them resolve both syntax and semantic errors. Given the extensive literature on this topic, we focus on a selection of representative studies most relevant to our work.

Leinonen et al.~utilized Codex, a model based on GPT-3, to improve the clarity of programming error messages \cite{Leinonen2023}. Programming error messages are often difficult for novice programmers to interpret \cite{Prather2018}. Their work aimed to enhance these messages by providing explanations of the errors along with suggested fixes. They found that LLMs can generate helpful, novice-friendly explanations, improving students' comprehension and their ability to correct errors. While the explanations generated by Codex were generally comprehensible, the suggested fixes were correct only 33\% of the time. To improve the accuracy of generated fixes, Phung et al.~introduced a novel runtime validation mechanism to assess whether the feedback provided by the LLM in the initial stage was suitable for students \cite{Phung2023}. Their approach involved iteratively querying the LLM to generate fixes for the buggy program while leveraging feedback from previous LLM iterations. If the number of syntactically correct programs exceeded a predefined threshold, the feedback was deemed acceptable. This method enabled them to achieve high precision in identifying and correcting syntax errors. However, their study was limited to syntax errors, and its effectiveness in addressing semantic errors has not been explored.

A widely explored approach for addressing semantic errors is automated program repair (APR) \cite{Zhang2023_tosem}. LLMs have been applied to APR through fine-tuning, few-shot learning and zero-shot learning \cite{Zhang2024b}. One notable example is PyDex, which employs few-shot learning to repair both syntactic and semantic bugs in introductory Python assignments \cite{Zhang2024}. PyDex utilizes the structure of a student's buggy program as input to the LLM, resulting in repairs that require fewer edits. When compared to other APR tools, PyDex achieved a repair rate of 96.5\%. Additionally, the average token edit distance for PyDex-generated patches was lower than that of competing tools. However, PyDex provides fixes without student intervention, which may hinder the development of debugging skills, as students receive corrections without actively engaging in the debugging process. CodeAid, on the other hand, provides helpful and technically accurate responses without revealing full code solutions to students \cite{Kazemitabaar2024}. This tool allows students to ask general programming questions, seek explanation for their code, request help in fixing their code, and receive help with writing new code. CodeAid employs an LLM in a two-step process to assist students in fixing their code. First, it generates a corrected version of the code based on the provided description and buggy implementation. Second, it explains the modifications using bullet points, detailing what was changed and why. However, the student interface only displays the bullet-point explanations rather than the corrected code itself, ensuring that solutions are not directly provided.

Although CodeAid does not directly provide solutions, it offers step-by-step guidance on necessary changes, potentially reducing the cognitive effort required for debugging. Carver and Risinger developed a traditional framework for training students in debugging \cite{Carver1987}. The process begins with testing a program, followed by answering a sequence of diagnostic questions, including ``what is the problem?'' and ``what type of bug could cause the problem?''. Katz and Anderson observed that programmers often employ backward reasoning when debugging their own code, typically starting by examining the output \cite{Katz1987}. Additionally, research indicates that the skills required for fixing errors are not necessarily related to the methods used to identify and locate them. Studies show that for most students, the main difficulty in debugging lies not in repairing errors, but in earlier troubleshooting stages, such as understanding the system, testing, and locating errors within it \cite{McCauley2008}. In fact, a multi-institutional study of novice debuggers by Fitzgerald et al.~found that once students successfully identify and locate bugs, they are generally able to fix them \cite{Fitzgerald2008}.

\section{Design of CodeHinter}

\begin{figure*}[t]
  \centering
  \begin{subfigure}{0.45\textwidth}
  \centering
  \includegraphics[height=0.26\textheight]{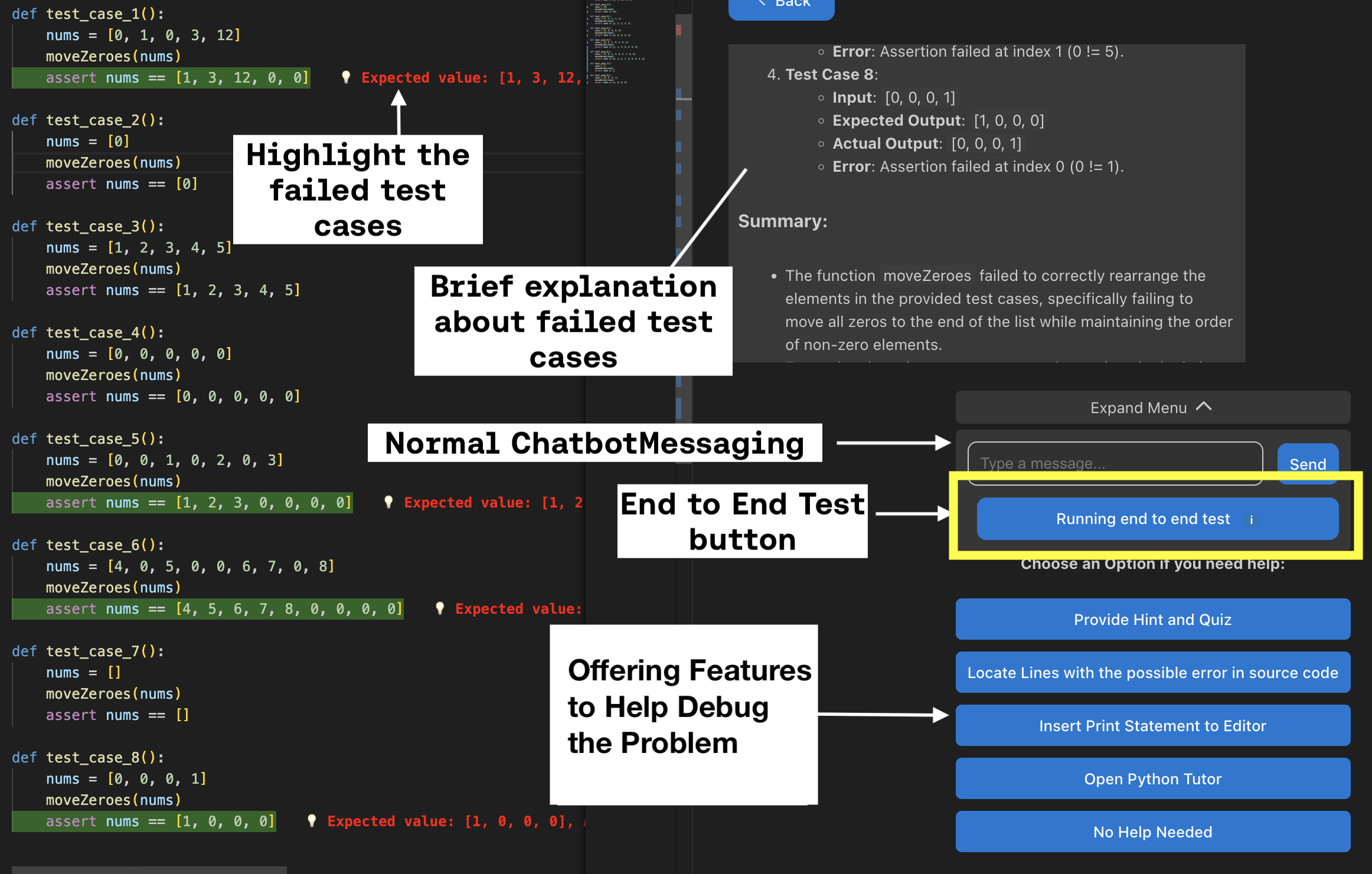}
  \end{subfigure}
  \begin{subfigure}{0.45\textwidth}
  \centering
  \includegraphics[height=0.26\textheight]{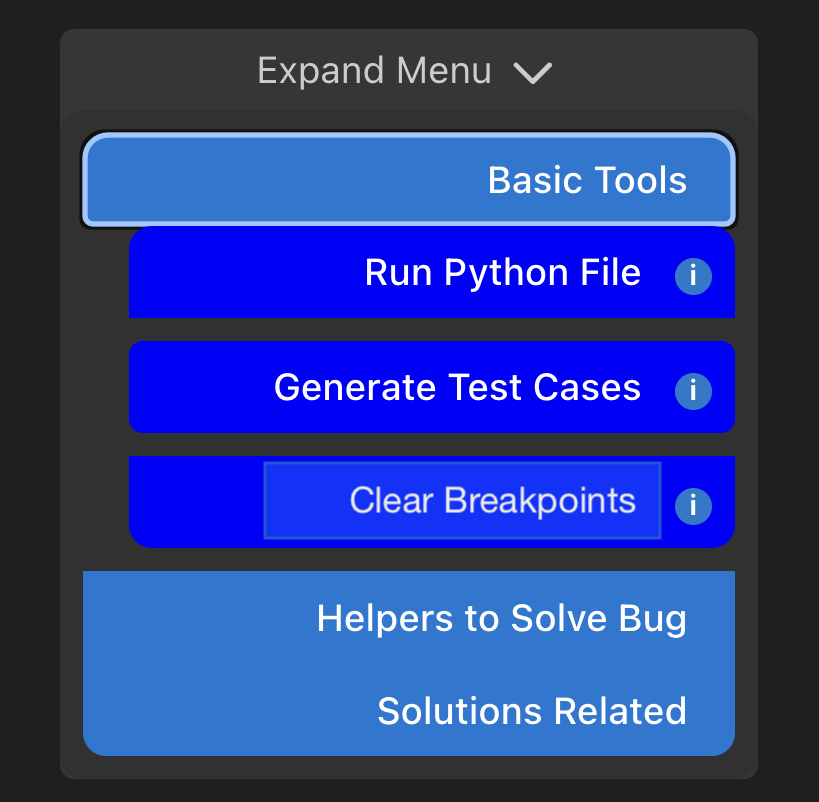}   
  \end{subfigure}
  \caption{\emph{(Left)} `End-to-End Test' feature, the main feature of \textsc{CodeHinter}. The screenshot shows the state when users encounter failed test cases (corresponding to * in Figure \ref{fig:user-flow}), highlighting the expected value and actual output in the text editor while providing a brief explanation in the chatbot. \emph{(Right)} \textsc{CodeHinter} `Expand Menu', where users can access other features.}
  \label{fig:main-screenshot}
\end{figure*}

Our tool, \textsc{CodeHinter}, shown in Figure \ref{fig:main-screenshot}, is designed to help students debug and fix their code. It evolved from an earlier version of the tool called \textsc{SID} (for Simulated Interactive Debugging)~\cite{noller2025sid}, which allowed users to run a test file for their Python programs. If any of the tests failed, SID automatically inserted breakpoints in the IDE. To achieve this functionality, the tool was integrated as an extension in Visual Studio Code (VS Code). An initial study on \textsc{SID} provided insights that informed further design improvements.

This paper presents the current version of \textsc{CodeHinter}, which continues to be developed as an extension within Visual Studio Code, following positive feedback from students. The key improvement over \textsc{SID} is a closer alignment with the debugging process to actively engage students in critical thinking. In particular, \textsc{CodeHinter} includes the following new features: (1) a single `End-to-End Test' button that allows students to run and test their code; (2) an LLM-powered system that generates hints and quizzes to prompt users to analyze errors and explore possible fixes; (3) line highlighting and a code difference interface to help students identify and modify relevant code sections; (4) a spectrum-based fault localization tool to pinpoint errors; and (5) an LLM-generated pseudo-code feature to help students understand the given problem.

Our tool currently supports Python, as it is one of the most widely used languages in introductory programming courses. Additionally, Python has several mature libraries for testing and fault localization that can be easily integrated. Our tool utilizes FauxPy~\cite{rezaalipour2024fauxpy} for spectrum-based fault localization, which is also part of the PyTest testing framework. The LLM powering our tool is OpenAI's GPT-4o, which offers high accuracy and significantly improved speed and cost-effectiveness compared to reasoning models like o1-mini and o1, which are better suited for more complex problems. GPT-4o is particularly effective for debugging and addressing novice programming issues~\cite{chatgpt-4o}.

\begin{figure}[t]
  \centering
  \includegraphics[width=\linewidth]{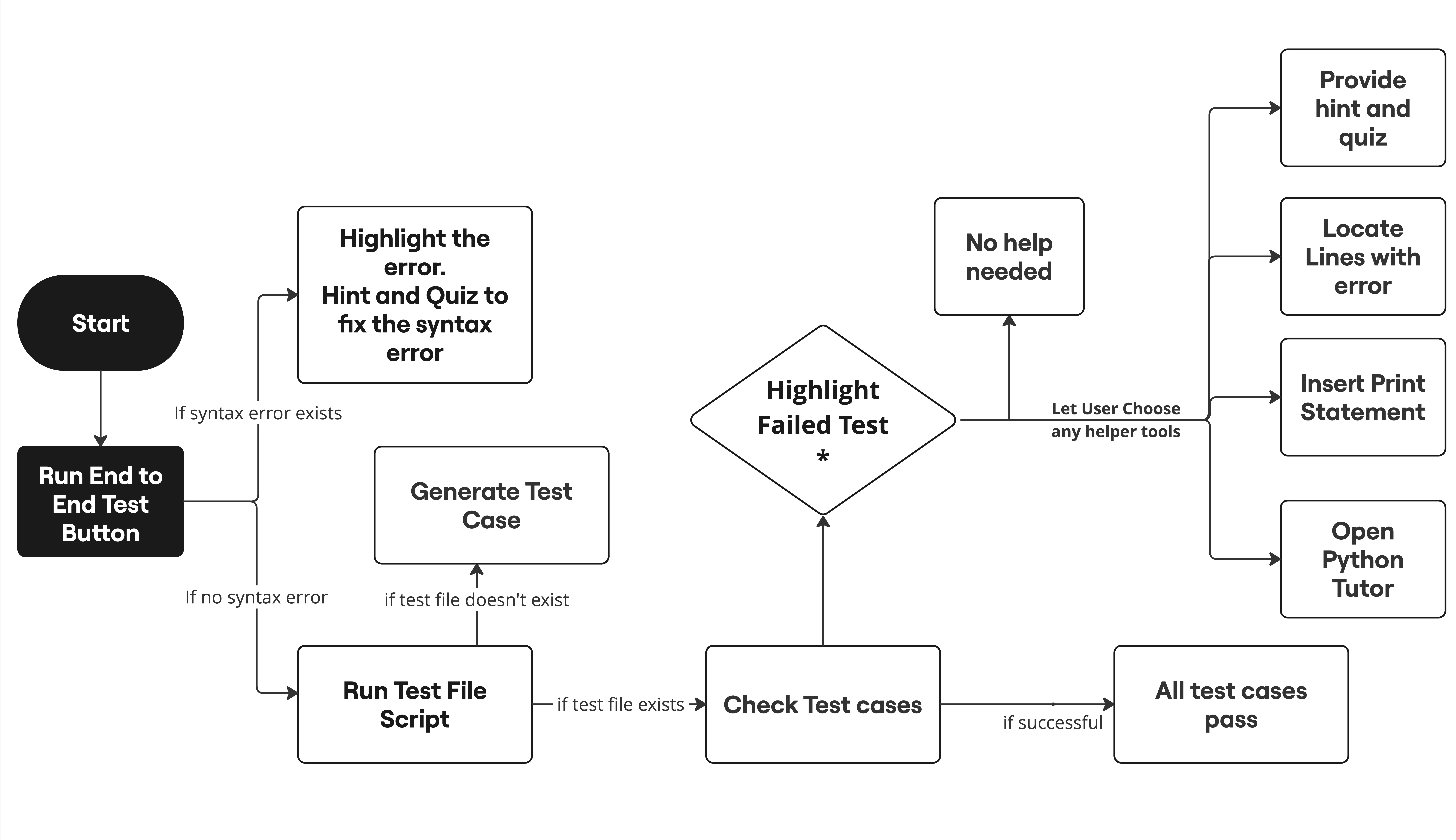}
  \caption{Flow of user interactions with our tool. The process begins with users pressing the `End-to-End Test' button, which checks for syntax and semantic errors. Users are then guided to complete the missing debugging steps. If test cases fail, the tool provides options to help users resolve the bugs. The asterisk indicates the state of the screenshot in Figure \ref{fig:main-screenshot}.}
  \label{fig:user-flow}
\end{figure}

One of the key features of the extension is the `End-to-End Test' button. This button is designed to align with the standard debugging framework, incorporating backward reasoning. Figure \ref{fig:user-flow} shows how the user is engaged at various debugging stages.

Note that we did not implement \textsc{SID}'s `insert breakpoints' feature in \textsc{CodeHinter}, as the new version of the tool utilizes print statements for debugging instead, which are hypothesized to be simpler 
for novices. In the pilot study, we ask participants to test both \textsc{CodeHinter} and \textsc{SID} to compare these alternative strategies.

The helper tools in the end-to-end tests can also be manually selected by using the main menu item `Helpers to Solve Bugs'. Four tools are currently provided: identifying lines with errors; providing hints and quizzes; inserting print statements; and opening Python Tutor~\cite{Guo2021}. These features are designed to assist users in debugging their code without directly providing solutions from an LLM. They can be triggered automatically when test cases fail, or manually by expanding the menu and selecting the desired option (Figure \ref{fig:main-screenshot}).
We expand on the helper tools below.

\textbf{Locate Lines With Errors.} When users select this option, the tool runs Python's spectrum-based fault localization method, FauxPy~\cite{rezaalipour2024fauxpy}. FauxPy generates probability scores for source code line numbers, indicating the most likely sources of failed test cases. The algorithm identifies and ranks multiple lines based on their likelihood of containing faults. From this ranked output, we extract up to three lines with the highest probabilities.

Next, we leverage an LLM to provide explanations for why these specific lines are the most probable sources of error. This approach enhances the objectivity of the debugging process by ensuring that LLM-generated insights are grounded in the output of fault localization rather than speculative reasoning. By doing so, we mitigate the risk of hallucination when identifying potential errors.

The tool then highlights the identified lines and provides explanations to help users fix the error. It is important to note that fault localization tools like FauxPy identify lines where variables are not updated correctly, but these may not always be the exact source of the error. For instance, if a bug originates from an incorrect condition in an if-else statement, leading to a wrong assignment, FauxPy can detect that the values are assigned incorrectly but will not highlight that the issue stems from the condition itself. Although this approach may seem unintuitive, identifying lines where the expected value differs from the actual executed value aligns with how programmers typically debug their code using backward reasoning. Once they identify discrepancies in values, they hypothesize potential causes and trace the issue back to its source.

\textbf{Provide Hint and Quiz.} This feature is activated either when a syntax error occurs or when a user selects this option after encountering a semantic error. Instead of directly providing an explanation, the LLM generates three possible solutions, with only one being correct. These options suggest possible ways to correct the error, whether syntactical or semantic. Once the user selects an answer, the system immediately indicates whether the choice was correct and provides an explanation. By incorporating interactive quizzes, this feature encourages users to think critically while still receiving structured support.

\textbf{Insert Print Statement.} Research has shown that novice programmers often rely on print statements rather than using debugger mode \cite{Fitzgerald2008}. To support this behavior, we leverage LLMs to suggest up to three key variables for users to observe by printing their values to the standard output. Once the LLM identifies the locations for the print statements, the tool provides a brief explanation in the chat about why these variables are relevant. A new tab is then opened, displaying the printed output with green-highlighted lines to help users visualize changes. Additionally, users have the option to paste the modified code into the text editor if they wish to incorporate the suggested changes.

One intentional design choice is that this process does not directly modify the source code, encouraging users to analyze the suggested change before making edits. The code is only modified when the user chooses to paste the suggested changes into the text editor. This approach allows users to manually insert print statements, selecting only the most relevant lines rather than applying all the suggested modifications. 

\textbf{Open Python Tutor.} Given that novices rarely use debugger mode, to bridge this gap, the final helper tool is a button that redirects users to Python Tutor, allowing them to visualize code execution in an interactive environment. This approach provides a more intuitive and accessible debugging experience without requiring users to step into the complexity of debugger mode. We see this as a stepping stone for users, helping them gradually transition to advanced debugging techniques commonly used by more experienced programmers.

\section{Methodology}

We conducted a study with ten participants, divided into two groups of five. Though this number seems small, it is considered enough for usability testing according to \cite{virzi1992}. Each participant was given 50 minutes to complete two tasks: (1) debugging a buggy program using \textsc{CodeHinter}; and (2) debugging another buggy program using the `insert breakpoints' feature of \textsc{SID}. The only difference between the two groups was the sequence in which they used the tools. One group used \textsc{CodeHinter} first, followed by SID, while the other group used SID first, followed by \textsc{CodeHinter}. Each buggy question contained a maximum of two incorrect lines of code, and corresponding test cases were provided for both debugging tasks.

Before the session began, participants completed a profiling questionnaire to assess their programming background. We also provided a demonstration of the tools. During the debugging tasks, users had the flexibility to choose which helper tools within \textsc{CodeHinter} to use based on their preferences. However, we encouraged them to avoid revealing the provided code solution unless they were unable to solve the problem after 15 minutes. The participants' chat sessions were stored in a MongoDB NoSQL database to track their interactions and usage patterns. This allowed us to monitor which tools they accessed during the study.

The two test questions were sourced from LeetCode \cite{LeetCodeMoveZeroes, LeetCodeSummaryRanges} and cover different algorithmic challenges. The first question, \textit{Move Zeroes} (\#283), requires participants to move all zeroes in an array to the end while maintaining the relative order of the non-zero elements. The second question, \textit{Summary Ranges} (\#228), asks participants to summarize a sorted array of numbers into concise ranges of consecutive elements. 

After completing the tasks, participants were required to submit a post-task survey and a standardized Brooke's system usability questionnaire \cite{brooke1996sus}. The post-task survey consisted of common questions applicable to both \textsc{CodeHinter} and \textsc{SID}, followed by tool-specific questions. The common section assessed general usability, effectiveness, and user preferences, while the tool-specific sections focused on unique features, user confidence in debugging, and areas for improvement. Additionally, participants provided qualitative feedback on what they liked, disliked, and suggested features for future enhancements. As a token of appreciation, participants received the equivalent of USD 20 upon completion of the study.

\section {Results}

\subsection{Profile of Participants}

We recruited eight first-year and two second-year undergraduate students from the Singapore University of Technology and Design (SUTD). The first-year students had completed an introductory programming course but had not yet chosen their major, while the second-year students may have taken additional programming courses. We found that seven out of ten participants had experience writing over 500 lines of code for unique projects, suggesting they were not complete beginners. However, their confidence levels in debugging varied. While 50\% felt confident in debugging simpler problems, only 40\% were comfortable handling more complex issues. Regarding debugging preferences, 60\% preferred guidance that included explanations of key actions needed to identify the bug. Notably, only two students regularly used debugging tools as part of their programming habits, while five students had never heard of any debugging tools.

\subsection{Post-Task Survey Results}

\begin{figure}[t]
  \centering
  \includegraphics[width=\linewidth]{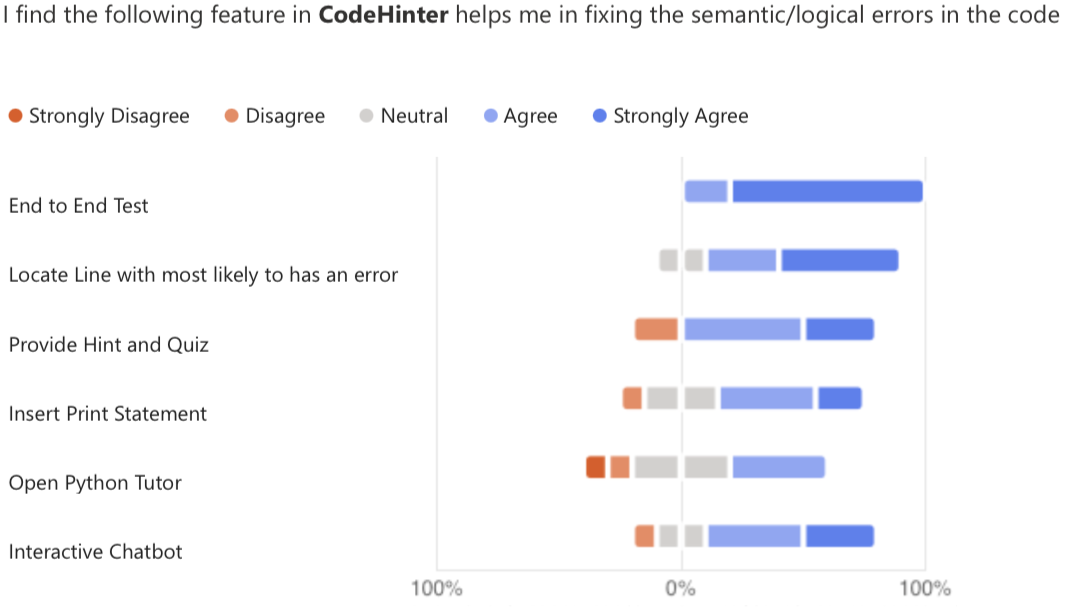}
  \caption{User perspectives on the usefulness of \textsc{CodeHinter}'s features for fixing semantic errors.}
  \label{fig:result-features}
  \Description{Codehinter.}
\end{figure}
We collected survey responses from participants after they completed all the assigned tasks. Due to limited space, we only highlight results for \textsc{CodeHinter} and not SID. Figure \ref{fig:result-features} illustrates the features that participants found most useful in \textsc{CodeHinter}. The results indicate that the most helpful tools were `End-to-End Test' and `Locate Lines With Errors', which aligns with our expectations.

\begin{figure}[t]
  \centering
  \includegraphics[width=\linewidth]{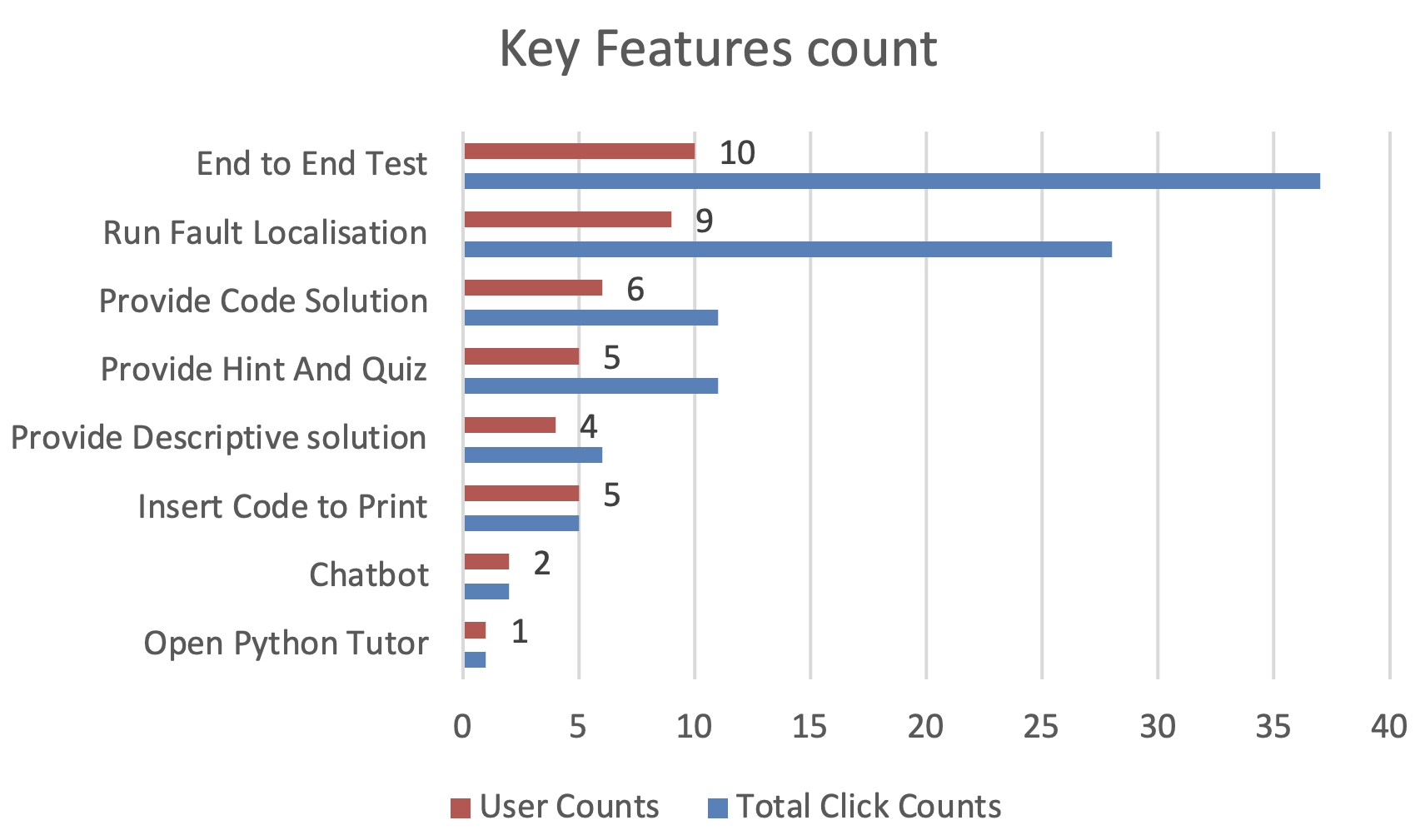}
  \caption{Frequency of access and utilization of features.}
  \label{fig:result-frequency}
\end{figure}
Additionally, we retrieved data from our database to track chatbot sessions and identify the most frequently used tools. The data shown in Figure~\ref{fig:result-frequency} only counts clicks on individual menu options, excluding instances where features were used as part of the `End-to-End Test' button.

As expected, `End-to-End Test' was used by all participants and ranked as the most frequently used feature. Among the four available helper tools, `Locate Lines With Errors' was the most popular, with most participants using it multiple times, except for one user. The `Provide Hint and Quiz' and `Insert Print Statements' features were each used by half of the participants. However, participants tended to use `Provide Hint and Quiz' multiple times, whereas `Insert Print Statements' was typically used only once per user. Additionally, only one user utilized the `Open Python Tutor in Web Browser' feature, while two users interacted with the general chatbot. Notably, six out of ten participants clicked the `Provide Code Solution' button, with some clicking it multiple times, possibly to check and compare their answers. 

\subsection{Usability and Satisfaction}

Using the standardized Brooke’s System Usability Scale (SUS) \cite{brooke1996sus}, we calculated an average score of 75/100 for \textsc{CodeHinter}, compared to 65/100 for SID, as reported in our first study~\cite{noller2025sid}. According to Brooke’s methodology, SUS scores are most meaningful when compared across systems, allowing us to conclude that \textsc{CodeHinter} provides a significantly improved user experience over SID. This finding is further reinforced by the post-task survey, where participants gave an overall satisfaction rating of 89/100 for \textsc{CodeHinter}, compared to 60/100 for SID. The higher rating for \textsc{CodeHinter} suggests that users found it more effective and user-friendly in assisting with debugging tasks, indicating a strong improvement in our second design iteration.


Participants highlighted the strengths of \textsc{CodeHinter}, particularly in debugging semantic errors and its ease of use. Selected user quotes from the survey include:

\begin{quote}
``It has a lot of features to help us debug semantic errors instead of simply copying and pasting solutions from other LLM AI models.''\\
``It was user-friendly and intuitive to use, providing accurate information.''\\
``I do not dislike it because of its high quality in helping beginners become more familiar with coding.''
\end{quote}

\subsubsection{End to End Test}

Users expressed the highest level of satisfaction with this feature. All participants found it helpful, with two selecting `Agree' and eight selecting `Strongly Agree' in response to the statement that it significantly aided in problem-solving. 

\begin{quote}
``It helps me to know which logic errors I am still facing and also display the failed test cases that I need more time to debug the semantic errors.'' \\
``It was very useful to have a tool that could provide me with descriptions of what happened during the testing of the test cases.''
\end{quote}

\subsubsection{Locate Line With Error}

Among the four helper tools, `Locate Lines With Errors' was the most frequently used and highly favored by users. In the survey measuring how often participants would use this feature, it received an average rating of 4.5/5. Interestingly, our database retrieval showed that one participant did not use this tool at all, as this participant was able to solve the problem just by utilizing `End to End Test'.

\begin{quote}
``This feature allowed me to more easily find possible logical flaws in my code which could've taken me a lot more time to spot without such tools.''\\
``This is by far the best feature which helps us seamlessly navigate where the bug is most likely to be.''\\
``I don't have to search and debug/print a million lines to find the root cause of the error.''
\end{quote}

\subsubsection{Hint and Quiz}
This feature was also well-received by users. The majority of the participants appreciated its interactive approach, as it encourages thinking when fixing the bugs. However, one comment suggested that this feature may not be their preferred option if given a limited time to solve the problems in. 

\begin{quote}
``The hints provided make me think about where the error is in my code by myself.''\\
``Effective. It is interactive and helps the user to understand the bug.''\\
``I am unlikely to use this tool unless I am very desperate to solve a problem within a time duration, and in that situation I wouldn't want to be quizzed with possible solutions.''
\end{quote}

\subsubsection{Insert Print Statement}

The results for `Insert Print Statement' are similar to the `Provide Hint and Quiz' feature, with half of the participants attempting to use it. However, each user only used this feature once, unlike `Provide Hint and Quiz', which participants interacted with multiple times. This difference may be because locations and variables of the print statements tend to be the same depending on the structure of the code. On the other hand, the `Provide Hint and Quiz' feature generates different options as the code changes. This could happen if the users apply a wrong fix. Users also mentioned that `Insert Print Statement' helps them debug faster by allowing them to quickly spot errors, making it especially beneficial for beginners.

\begin{quote}
``I think this tool can be useful for simple testing and spotting errors in the code, especially for beginners who are less familiar with the debugging tool.''\\
``Can do the print statements for me faster, which normally takes a while to do.''
\end{quote}

\section{Discussion \& Future Work}

Overall, participants provided positive feedback on \textsc{CodeHinter}. As expected, users found the `End-to-End Test' button highly useful, as indicated by both its frequent usage among all participants and their comments about the tool. Among the individual tools designed to assist with debugging, users found the `Locate Lines With Errors' feature the most helpful. This aligns with previous studies showing that one of the most challenging aspects of debugging for novice programmers is locating errors~\cite{McCauley2008, Fitzgerald2008}. Additionally, research on the use of APR has similarly found that error location messages are the most helpful information for novice programmers when fixing their code~\cite{kurniawanclara}.

Our study results also showed positive feedback on the use of the `Hints and Quiz' helper tool to engage the participants in the debugging process. Instead of directly specifying what to replace, as most previous tools do, incorporating quizzes encourages learners to think critically and formulate hypotheses about the possible causes of errors and potential solutions. Participants' comments after using the tool aligned with our intended approach, confirming that novice debuggers can be actively involved in aspects of the debugging process, such as hypothesizing the cause of errors and identifying possible fixes.

Regarding the `Insert Print Statement' feature, we were initially surprised that each user used it only once for the problem they worked on with \textsc{CodeHinter}. However, their comments were largely positive, indicating that they found the feature useful. Therefore, we hypothesize that, unlike `Provide Hints and Quiz', this feature serves a one-time function in helping users identify bugs. Once the print statements are inserted, users do not need to reinsert them; instead, they simply run the code to observe the output.
Additionally, this feature was introduced as part of our design exploration, serving as an alternative to automatically inserting breakpoints, as implemented in the first iteration of our tool, \textsc{SID}. Our initial hypothesis was that novice programmers find using print statements easier than working with an IDE's debug mode. However, we discovered that some intermediate programmers prefer using breakpoints and the debug mode instead. This suggests that future iterations of the tool should offer both options, allowing users to choose the method that best suits their debugging preferences.

This leads us to conclude that personalizing the tool based on the user's profile is essential. With advancements in AI, it is increasingly feasible to profile users based on their interactions with debugging tools. By doing so, novice programmers can be prompted with features better suited to their skill level, while more experienced users can be provided with advanced debugging tools tailored to their needs. Additionally, the tool can be adapted based on the specific problem being debugged, ensuring a more effective and user-centric debugging experience.

Furthermore, the participants also provided valuable suggestions for future improvements. For example, participants suggested incorporating real-time assistance not only for writing code but also for debugging, enabling dynamic support as they work through identifying and fixing bugs. Additionally, they expressed interest in integrating code quality analysis during debugging to provide more tailored feedback.

This work is limited by its small participant pool, as it represents a pilot study focused on the tool's design and usability. Further research should examine its impact on learning outcomes,  particularly how it influences the development of debugging skills among novice programmers. 

\section{Conclusion}
In conclusion, this study demonstrates the potential of AI-assisted tools like \textsc{CodeHinter} to enhance the debugging experience for novice programmers by fostering active engagement rather than passive reliance on generating complete solutions. Through a user-centric design that integrates spectrum-based fault localization, interactive hints and quizzes, and print statement suggestions, all within a familiar IDE environment, \textsc{CodeHinter} effectively supports the debugging process. The tool’s usability and effectiveness were validated in a pilot study, with participants favoring features that guided them through the problem-solving process while reinforcing critical thinking. Our findings also underscore the importance of tailoring AI support to individual user profiles and problem contexts. Future debugging tools should focus on providing personalized assistance to optimize learning outcomes and foster independent debugging skills.

\begin{acks}
This work was supported by the Ministry of Education, Singapore, under the Tertiary Research Fund (MOE-TRF) Grant No. MOE2023-TRF-034. We gratefully acknowledge this support, which made this research possible. We would also like to acknowledge the use of OpenAI’s ChatGPT for assisting in the refinement of English and improvement of paragraph clarity throughout the writing process.
This study was reviewed and approved by the Institutional Review Board (IRB) of the Singapore University of Technology and Design (SUTD) under IRB protocol number S-24-669. All participants provided informed consent prior to their involvement in the study.
\end{acks}

\bibliographystyle{ACM-Reference-Format}
\bibliography{references}


\begin{thebibliography}{29}


\ifx \showCODEN    \undefined \def \showCODEN     #1{\unskip}     \fi
\ifx \showISBNx    \undefined \def \showISBNx     #1{\unskip}     \fi
\ifx \showISBNxiii \undefined \def \showISBNxiii  #1{\unskip}     \fi
\ifx \showISSN     \undefined \def \showISSN      #1{\unskip}     \fi
\ifx \showLCCN     \undefined \def \showLCCN      #1{\unskip}     \fi
\ifx \shownote     \undefined \def \shownote      #1{#1}          \fi
\ifx \showarticletitle \undefined \def \showarticletitle #1{#1}   \fi
\ifx \showURL      \undefined \def \showURL       {\relax}        \fi
\providecommand\bibfield[2]{#2}
\providecommand\bibinfo[2]{#2}
\providecommand\natexlab[1]{#1}
\providecommand\showeprint[2][]{arXiv:#2}

\bibitem[Becker et~al\mbox{.}(2019)]%
        {Becker2019}
\bibfield{author}{\bibinfo{person}{Brett~A Becker}, \bibinfo{person}{Paul
  Denny}, \bibinfo{person}{Raymond Pettit}, \bibinfo{person}{Durell Bouchard},
  \bibinfo{person}{Dennis~J Bouvier}, \bibinfo{person}{Brian Harrington},
  \bibinfo{person}{Amir Kamil}, \bibinfo{person}{Amey Karkare},
  \bibinfo{person}{Chris McDonald}, \bibinfo{person}{Peter-Michael Osera},
  \bibinfo{person}{Janice~L Pearce}, {and} \bibinfo{person}{James Prather}.}
  \bibinfo{year}{2019}\natexlab{}.
\newblock \showarticletitle{Compiler Error Messages Considered Unhelpful: The
  Landscape of Text-Based Programming Error Message Research}. In
  \bibinfo{booktitle}{\emph{Proceedings of the Working Group Reports on
  Innovation and Technology in Computer Science Education}} (New York, NY,
  USA). \bibinfo{publisher}{Association for Computing Machinery},
  \bibinfo{pages}{177--210}.
\newblock
\showISBNx{9781450375672}


\bibitem[Brooke(1996)]%
        {brooke1996sus}
\bibfield{author}{\bibinfo{person}{John Brooke}.}
  \bibinfo{year}{1996}\natexlab{}.
\newblock \showarticletitle{SUS: A quick and dirty usability scale}.
\newblock \bibinfo{journal}{\emph{Usability Evaluation in Industry}}
  (\bibinfo{year}{1996}).
\newblock


\bibitem[Campbell et~al\mbox{.}(2014)]%
        {Campbell2014}
\bibfield{author}{\bibinfo{person}{Joshua~Charles Campbell},
  \bibinfo{person}{Abram Hindle}, {and} \bibinfo{person}{José~Nelson Amaral}.}
  \bibinfo{year}{2014}\natexlab{}.
\newblock \showarticletitle{Syntax errors just aren't natural: improving error
  reporting with language models}. In \bibinfo{booktitle}{\emph{Proceedings of
  the 11th Working Conference on Mining Software Repositories}} (New York, NY,
  USA). \bibinfo{publisher}{Association for Computing Machinery},
  \bibinfo{pages}{252--261}.
\newblock
\showISBNx{9781450328630}


\bibitem[Carver and Risinger(1987)]%
        {Carver1987}
\bibfield{author}{\bibinfo{person}{Sharon~McCoy Carver} {and}
  \bibinfo{person}{Sally~Clarke Risinger}.} \bibinfo{year}{1987}\natexlab{}.
\newblock \bibinfo{booktitle}{\emph{Improving children's debugging skills}}.
\newblock \bibinfo{publisher}{Ablex Publishing Corp.},
  \bibinfo{pages}{147--171}.
\newblock
\showISBNx{0893914614}


\bibitem[Collins(2023)]%
        {Collins2023}
\bibfield{author}{\bibinfo{person}{Jonathan~E Collins}.}
  \bibinfo{year}{2023}\natexlab{}.
\newblock \showarticletitle{Policy Solutions: Policy questions for ChatGPT and
  artificial intelligence}.
\newblock \bibinfo{journal}{\emph{Phi Delta Kappan}}  \bibinfo{volume}{104}
  (\bibinfo{year}{2023}), \bibinfo{pages}{60--61}.
\newblock
Issue 7.


\bibitem[Denny et~al\mbox{.}(2012)]%
        {Denny2012}
\bibfield{author}{\bibinfo{person}{Paul Denny}, \bibinfo{person}{Andrew
  Luxton-Reilly}, {and} \bibinfo{person}{Ewan Tempero}.}
  \bibinfo{year}{2012}\natexlab{}.
\newblock \showarticletitle{All syntax errors are not equal}. In
  \bibinfo{booktitle}{\emph{Proceedings of the 17th ACM Annual Conference on
  Innovation and Technology in Computer Science Education}} (New York, NY,
  USA). \bibinfo{publisher}{Association for Computing Machinery},
  \bibinfo{pages}{75--80}.
\newblock
\showISBNx{9781450312462}


\bibitem[Fitzgerald et~al\mbox{.}(2008)]%
        {Fitzgerald2008}
\bibfield{author}{\bibinfo{person}{Sue Fitzgerald}, \bibinfo{person}{Gary
  Lewandowski}, \bibinfo{person}{Renée McCauley}, \bibinfo{person}{Laurie
  Murphy}, \bibinfo{person}{Beth Simon}, \bibinfo{person}{Lynda Thomas}, {and}
  \bibinfo{person}{Carol Zander}.} \bibinfo{year}{2008}\natexlab{}.
\newblock \showarticletitle{Debugging: Finding, fixing and flailing, a
  multi-institutional study of novice debuggers}.
\newblock \bibinfo{journal}{\emph{Computer Science Education}}
  \bibinfo{volume}{18} (\bibinfo{year}{2008}), \bibinfo{pages}{93--116}.
\newblock
Issue 2.
\showISSN{17445175}


\bibitem[GitHub(2023)]%
        {copilot}
\bibfield{author}{\bibinfo{person}{GitHub}.} \bibinfo{year}{2023}\natexlab{}.
\newblock \bibinfo{title}{GitHub Copilot}.
\newblock
\urldef\tempurl%
\url{https://github.com/features/copilot}
\showURL{%
\tempurl}
\newblock
\shownote{Accessed: March 12, 2025}.


\bibitem[Guo(2021)]%
        {Guo2021}
\bibfield{author}{\bibinfo{person}{Philip Guo}.}
  \bibinfo{year}{2021}\natexlab{}.
\newblock \showarticletitle{Ten Million Users and Ten Years Later: Python
  Tutor's Design Guidelines for Building Scalable and Sustainable Research
  Software in Academia}. In \bibinfo{booktitle}{\emph{UIST 2021 - Proceedings
  of the 34th Annual ACM Symposium on User Interface Software and Technology}}.
  \bibinfo{publisher}{Association for Computing Machinery, Inc},
  \bibinfo{pages}{1235--1251}.
\newblock
\showISBNx{9781450386357}


\bibitem[Joshi et~al\mbox{.}(2023)]%
        {Joshi2023}
\bibfield{author}{\bibinfo{person}{Harshit Joshi},
  \bibinfo{person}{José~Cambronero Sanchez}, \bibinfo{person}{Sumit Gulwani},
  \bibinfo{person}{Vu Le}, \bibinfo{person}{Gust Verbruggen}, {and}
  \bibinfo{person}{Ivan Radiček}.} \bibinfo{year}{2023}\natexlab{}.
\newblock \showarticletitle{Repair Is Nearly Generation: Multilingual Program
  Repair with LLMs}.
\newblock \bibinfo{journal}{\emph{Proceedings of the AAAI Conference on
  Artificial Intelligence}}  \bibinfo{volume}{37} (\bibinfo{date}{6}
  \bibinfo{year}{2023}), \bibinfo{pages}{5131--5140}.
\newblock
Issue 4.
\showISSN{2374-3468}


\bibitem[Katz and Anderson(1987)]%
        {Katz1987}
\bibfield{author}{\bibinfo{person}{Irvin~R Katz} {and} \bibinfo{person}{John~R
  Anderson}.} \bibinfo{year}{1987}\natexlab{}.
\newblock \showarticletitle{Debugging: an analysis of bug-location strategies}.
\newblock \bibinfo{journal}{\emph{Hum.-Comput. Interact.}}  \bibinfo{volume}{3}
  (\bibinfo{date}{12} \bibinfo{year}{1987}), \bibinfo{pages}{351--399}.
\newblock
Issue 4.
\showISSN{0737-0024}


\bibitem[Kazemitabaar et~al\mbox{.}(2024)]%
        {Kazemitabaar2024}
\bibfield{author}{\bibinfo{person}{Majeed Kazemitabaar},
  \bibinfo{person}{Runlong Ye}, \bibinfo{person}{Xiaoning Wang},
  \bibinfo{person}{Austin~Z. Henley}, \bibinfo{person}{Paul Denny},
  \bibinfo{person}{Michelle Craig}, {and} \bibinfo{person}{Tovi Grossman}.}
  \bibinfo{year}{2024}\natexlab{}.
\newblock \showarticletitle{CodeAid: Evaluating a Classroom Deployment of an
  LLM-based Programming Assistant that Balances Student and Educator Needs}. In
  \bibinfo{booktitle}{\emph{Conference on Human Factors in Computing Systems -
  Proceedings}}. \bibinfo{publisher}{Association for Computing Machinery}.
\newblock
\showISBNx{9798400703300}


\bibitem[Kurniawan et~al\mbox{.}(2023)]%
        {kurniawanclara}
\bibfield{author}{\bibinfo{person}{Oka Kurniawan},
  \bibinfo{person}{Christopher~M. Poskitt}, \bibinfo{person}{Ismam Al~Hoque},
  \bibinfo{person}{Norman Tiong~Seng Lee}, \bibinfo{person}{Cyrille Jégourel},
  {and} \bibinfo{person}{Nachamma Sockalingam}.}
  \bibinfo{year}{2023}\natexlab{}.
\newblock \showarticletitle{How Helpful do Novice Programmers Find the Feedback
  of an Automated Repair Tool?}. In \bibinfo{booktitle}{\emph{2023 IEEE
  International Conference on Teaching, Assessment and Learning for Engineering
  (TALE)}}. \bibinfo{pages}{1--6}.
\newblock


\bibitem[LeetCode(nda)]%
        {LeetCodeMoveZeroes}
\bibfield{author}{\bibinfo{person}{LeetCode}.}
  \bibinfo{year}{n.d.}\natexlab{a}.
\newblock \bibinfo{title}{Move Zeroes}.
\newblock
\urldef\tempurl%
\url{https://leetcode.com/problems/move-zeroes/description/}
\showURL{%
\tempurl}
\newblock
\shownote{Accessed: March 12, 2025}.


\bibitem[LeetCode(ndb)]%
        {LeetCodeSummaryRanges}
\bibfield{author}{\bibinfo{person}{LeetCode}.}
  \bibinfo{year}{n.d.}\natexlab{b}.
\newblock \bibinfo{title}{Summary Ranges}.
\newblock
\urldef\tempurl%
\url{https://leetcode.com/problems/summary-ranges/description/}
\showURL{%
\tempurl}
\newblock
\shownote{Accessed: March 12, 2025}.


\bibitem[Leinonen et~al\mbox{.}(2023)]%
        {Leinonen2023}
\bibfield{author}{\bibinfo{person}{Juho Leinonen}, \bibinfo{person}{Arto
  Hellas}, \bibinfo{person}{Sami Sarsa}, \bibinfo{person}{Brent Reeves},
  \bibinfo{person}{Paul Denny}, \bibinfo{person}{James Prather}, {and}
  \bibinfo{person}{Brett~A. Becker}.} \bibinfo{year}{2023}\natexlab{}.
\newblock \showarticletitle{Using Large Language Models to Enhance Programming
  Error Messages}. In \bibinfo{booktitle}{\emph{SIGCSE 2023 - Proceedings of
  the 54th ACM Technical Symposium on Computer Science Education}},
  Vol.~\bibinfo{volume}{1}. \bibinfo{publisher}{Association for Computing
  Machinery, Inc}, \bibinfo{pages}{563--569}.
\newblock
\showISBNx{9781450394314}


\bibitem[McCauley et~al\mbox{.}(2008)]%
        {McCauley2008}
\bibfield{author}{\bibinfo{person}{Renée McCauley}, \bibinfo{person}{Sue
  Fitzgerald}, \bibinfo{person}{Gary Lewandowski}, \bibinfo{person}{Laurie
  Murphy}, \bibinfo{person}{Beth Simon}, \bibinfo{person}{Lynda Thomas}, {and}
  \bibinfo{person}{Carol Zander}.} \bibinfo{year}{2008}\natexlab{}.
\newblock \bibinfo{title}{Debugging: A review of the literature from an
  educational perspective}.
\newblock
Issue 2.
\showISSN{17445175}


\bibitem[Noller et~al\mbox{.}(2025)]%
        {noller2025sid}
\bibfield{author}{\bibinfo{person}{Yannic Noller}, \bibinfo{person}{Erick
  Chandra}, \bibinfo{person}{Srinidhi Chandrashekar}, \bibinfo{person}{Kenny
  Choo}, \bibinfo{person}{Cyrille Jegourel}, \bibinfo{person}{Oka Kurniawan},
  {and} \bibinfo{person}{Christopher~M. Poskitt}.}
  \bibinfo{year}{2025}\natexlab{}.
\newblock \showarticletitle{Simulated Interactive Debugging}. In
  \bibinfo{booktitle}{\emph{Proc.\ IEEE/ACM International Conference on
  Automated Software Engineering (ASE'25): New Ideas and Emerging Results}}.
  \bibinfo{publisher}{ACM}.
\newblock


\bibitem[OpenAI(2022)]%
        {chatgpt}
\bibfield{author}{\bibinfo{person}{OpenAI}.} \bibinfo{year}{2022}\natexlab{}.
\newblock \bibinfo{title}{Introducing ChatGPT}.
\newblock
\urldef\tempurl%
\url{https://openai.com/index/chatgpt/}
\showURL{%
\tempurl}
\newblock
\shownote{Accessed: March 12, 2025}.


\bibitem[OpenAI(2024)]%
        {chatgpt-4o}
\bibfield{author}{\bibinfo{person}{OpenAI}.} \bibinfo{year}{2024}\natexlab{}.
\newblock \bibinfo{title}{Learning to reason with LLMs}.
\newblock
\urldef\tempurl%
\url{https://openai.com/index/learning-to-reason-with-llms/}
\showURL{%
\tempurl}
\newblock
\shownote{Accessed: March 12, 2025}.


\bibitem[Phung et~al\mbox{.}(2023)]%
        {Phung2023}
\bibfield{author}{\bibinfo{person}{Tung Phung}, \bibinfo{person}{Jos{\'{e}}
  Cambronero}, \bibinfo{person}{Sumit Gulwani}, \bibinfo{person}{Tobias Kohn},
  \bibinfo{person}{Rupak Majumdar}, \bibinfo{person}{Adish Singla}, {and}
  \bibinfo{person}{Gustavo Soares}.} \bibinfo{year}{2023}\natexlab{}.
\newblock \showarticletitle{Generating High-Precision Feedback for Programming
  Syntax Errors using Large Language Models}. In
  \bibinfo{booktitle}{\emph{Proceedings of the 16th International Conference on
  Educational Data Mining, {EDM} 2023, Bengaluru, India, July 11-14, 2023}}.
  \bibinfo{publisher}{International Educational Data Mining Society}.
\newblock


\bibitem[Prather et~al\mbox{.}(2018)]%
        {Prather2018}
\bibfield{author}{\bibinfo{person}{James Prather}, \bibinfo{person}{Raymond
  Pettit}, \bibinfo{person}{Kayla McMurry}, \bibinfo{person}{Alani Peters},
  \bibinfo{person}{John Homer}, {and} \bibinfo{person}{Maxine Cohen}.}
  \bibinfo{year}{2018}\natexlab{}.
\newblock \showarticletitle{Metacognitive Difficulties Faced by Novice
  Programmers in Automated Assessment Tools}. In
  \bibinfo{booktitle}{\emph{Proceedings of the 2018 ACM Conference on
  International Computing Education Research}} (New York, NY, USA).
  \bibinfo{publisher}{Association for Computing Machinery},
  \bibinfo{pages}{41--50}.
\newblock
\showISBNx{9781450356282}


\bibitem[Rezaalipour and Furia(2024)]%
        {rezaalipour2024fauxpy}
\bibfield{author}{\bibinfo{person}{Mohammad Rezaalipour} {and}
  \bibinfo{person}{Carlo~A Furia}.} \bibinfo{year}{2024}\natexlab{}.
\newblock \showarticletitle{FauxPy: A Fault Localization Tool for Python}.
\newblock \bibinfo{journal}{\emph{arXiv preprint arXiv:2404.18596}}
  (\bibinfo{year}{2024}).
\newblock


\bibitem[Smith and Rixner(2019)]%
        {Smith2019}
\bibfield{author}{\bibinfo{person}{Rebecca Smith} {and} \bibinfo{person}{Scott
  Rixner}.} \bibinfo{year}{2019}\natexlab{}.
\newblock \showarticletitle{The error landscape: Characterizing the mistakes of
  novice programmers}. In \bibinfo{booktitle}{\emph{SIGCSE 2019 - Proceedings
  of the 50th ACM Technical Symposium on Computer Science Education}}.
  \bibinfo{publisher}{Association for Computing Machinery, Inc},
  \bibinfo{pages}{538--544}.
\newblock
\showISBNx{9781450358903}


\bibitem[Virzi(1992)]%
        {virzi1992}
\bibfield{author}{\bibinfo{person}{Robert~A. Virzi}.}
  \bibinfo{year}{1992}\natexlab{}.
\newblock \showarticletitle{Refining the Test Phase of Usability Evaluation:
  How Many Subjects Is Enough?}
\newblock \bibinfo{journal}{\emph{Human Factors}} \bibinfo{volume}{34},
  \bibinfo{number}{4} (\bibinfo{year}{1992}), \bibinfo{pages}{457--468}.
\newblock


\bibitem[Zhang et~al\mbox{.}(2024a)]%
        {Zhang2024}
\bibfield{author}{\bibinfo{person}{Jialu Zhang}, \bibinfo{person}{José~Pablo
  Cambronero}, \bibinfo{person}{Sumit Gulwani}, \bibinfo{person}{Vu Le},
  \bibinfo{person}{Ruzica Piskac}, \bibinfo{person}{Gustavo Soares}, {and}
  \bibinfo{person}{Gust Verbruggen}.} \bibinfo{year}{2024}\natexlab{a}.
\newblock \showarticletitle{PyDex: Repairing Bugs in Introductory Python
  Assignments using LLMs}.
\newblock \bibinfo{journal}{\emph{Proceedings of the ACM on Programming
  Languages}}  \bibinfo{volume}{8} (\bibinfo{date}{4} \bibinfo{year}{2024}).
\newblock
Issue OOPSLA1.
\showISSN{24751421}


\bibitem[Zhang et~al\mbox{.}(2023a)]%
        {Zhang2023_tosem}
\bibfield{author}{\bibinfo{person}{Quanjun Zhang}, \bibinfo{person}{Chunrong
  Fang}, \bibinfo{person}{Yuxiang Ma}, \bibinfo{person}{Weisong Sun}, {and}
  \bibinfo{person}{Zhenyu Chen}.} \bibinfo{year}{2023}\natexlab{a}.
\newblock \showarticletitle{A Survey of Learning-based Automated Program
  Repair}.
\newblock \bibinfo{journal}{\emph{ACM Trans. Softw. Eng. Methodol.}}
  \bibinfo{volume}{33} (\bibinfo{date}{12} \bibinfo{year}{2023}).
\newblock
Issue 2.
\showISSN{1049-331X}


\bibitem[Zhang et~al\mbox{.}(2024b)]%
        {Zhang2024b}
\bibfield{author}{\bibinfo{person}{Quanjun Zhang}, \bibinfo{person}{Chunrong
  Fang}, \bibinfo{person}{Yang Xie}, \bibinfo{person}{YuXiang Ma},
  \bibinfo{person}{Weisong Sun}, \bibinfo{person}{Yun Yang}, {and}
  \bibinfo{person}{Zhenyu Chen}.} \bibinfo{year}{2024}\natexlab{b}.
\newblock \showarticletitle{A Systematic Literature Review on Large Language
  Models for Automated Program Repair}.
\newblock  (\bibinfo{date}{5} \bibinfo{year}{2024}).
\newblock
\urldef\tempurl%
\url{http://arxiv.org/abs/2405.01466}
\showURL{%
\tempurl}


\bibitem[Zhang et~al\mbox{.}(2023b)]%
        {Zhang2023}
\bibfield{author}{\bibinfo{person}{Quanjun Zhang}, \bibinfo{person}{Tongke
  Zhang}, \bibinfo{person}{Juan Zhai}, \bibinfo{person}{Chunrong Fang},
  \bibinfo{person}{Bowen Yu}, \bibinfo{person}{Weisong Sun}, {and}
  \bibinfo{person}{Zhenyu Chen}.} \bibinfo{year}{2023}\natexlab{b}.
\newblock \showarticletitle{A Critical Review of Large Language Model on
  Software Engineering: An Example from ChatGPT and Automated Program Repair}.
\newblock  (\bibinfo{date}{10} \bibinfo{year}{2023}).
\newblock
\urldef\tempurl%
\url{http://arxiv.org/abs/2310.08879}
\showURL{%
\tempurl}


\end{thebibliography}

\end{document}